# Real Time Topological Image Smoothing On Shared Memory Parallel Machines

Ramzi MAHMOUDI[1], Mohamed AKIL[1]

[1]*IGM, Unité Mixte CNRS-UMLV-ESIEE UMR8049, University Paris-Est*
*Cité Descartes, BP99, 93162 Noisy Le Grand, France*
*{mahmoudr, akilm}@esiee.fr*

**Abstract**

*Smoothing filter is the method of choice for image preprocessing and pattern recognition. We present a new concurrent method for smoothing 2D object in binary case. Proposed method provides a parallel computation while preserving the topology by using homotopic transformations. We introduce an adapted parallelization strategy called split, distribute and merge (SDM) strategy which allows efficient parallelization of a large class of topological operators including, mainly, smoothing, skeletonization, and watershed algorithms. To achieve a good speedup, we cared about task scheduling. Distributed work during smoothing process is done by a variable number of threads. Tests on 2D binary image (512\*512), using shared memory parallel machine (SMPM) with 8 CPU cores (2× Xeon E5405 running at frequency of 2 GHz), showed an enhancement of 5.2 thus a cadency of 32 images per second is achieved.*

## 1  Introduction

Smoothing is a fundamental step to reduce noise and prepare the image for subsequent processing such as segmentation. For example, the analysis or recognition of a shape is often perturbed by noise, thus the smoothing of object boundaries is a necessary preprocessing step. The smoothing procedure can also be used to extract some shape characteristics: by making the difference between the original and the smoothed object, salient or carved parts can be detected and measured. Smoothing shape has been extensively studied and many approaches have been proposed. The most popular one is the linear filtering by Laplacien smoothing for 2D-vector [1] and 3D mesh [2]. Other approach by morphological filtering can be applied directly to the shape [3] or to the curvature plot of the object's contour [4]. Unfortunately none of these operators preserve the topology (number of connected components) of the original image. In 2004, our team introduces a new method for smoothing 2D and 3D objects in binary images while preserving topology [5]. Objects are defined as sets of grid points and topology preservation is ensured by the exclusive use of homotopic transformations defined in the framework of digital topology [6]. Smoothness is obtained by the use of morphological openings and closings by metric discs or balls of increasing radius, in the manner of alternating sequential filters [7]. The authors' efforts have brought about two major issues such as preservation of the topology and the multitude of objects in the scene to smooth out without worrying about the latency of their filter.

This paper describes a new parallel method for topological smoothing which provides real time image processing. We present also an adapted parallelization strategy, called Split Distribute and Merge (SD&M). Our strategy is designed specifically for topological operators parallelizing on shared memory architectures. The new strategy is based upon the excusive combination of two patterns: divide and conquer and event-based coordination.

This paper is organized as follows: in section 2, some basic notions of topological operators are summarized; the original smoothing filter is introduced. In section 3, parallelization strategy, that has been adopted, is introduced. We define the class of operators that our parallelization strategy may cover. In section 4, concurrent computation method of topological smoothing is presented. Experimental analyzes results of different implementations are also presented and discussed. Finally, we conclude with summary and future work in section 5.

## 2  Theoretical background

In this section, we recall some basic notions of digital topology [6] and mathematical morphology for binary images [8]. We define also the homotopic alternating sequential filters [5]. For the sake of simplicity, we restrict ourselves to the minimal set of notions that will be useful for our purpose. We start by introducing morphological operators based on structuring elements which are balls in the sense of the Euclidean distance, in order to obtain the desired smoothing effect.

We denote by $\mathbb{Z}$ the set of relative integers, and by $E$ the discrete plane $\mathbb{Z}^2$. A point $x \in E$ is defined by $(x_1, x_2)$ with $x_i \in \mathbb{Z}$. Let $x \in E$ and $r \in \mathbb{N}$, we denote by $B_r(x)$ the ball of radius $r$ centred on $x$. $B_r(x)$ is defined by $\{y \in E, d(x,y) \leq r\}$, where $d$ is a distance on $E$. We denote by $B_r$ the map which associates to each $x$ in $E$ the ball $B_r(x)$. The Euclidean distance $d$ on $E$ is defined by:
$$d(x,y) = \left[ (x_1 - y_1)^2 - (x_2 - y_2)^2 \right]^{1/2}.$$

An operator in $E$ is a mapping from $P(E)$ into $P(E)$, where $P(E)$ denotes the set of all subsets $(\subset E)$. Let $r$ be an integer, the dilation $(\delta_r)$ by $B_r$ is defined by $\delta_r(X) = \bigcup_{x \in X} B_r(x)$, $\forall X \in P(E)$. The ball $B_r$ is termed as the structuring element of the dilation. The erosion $(\varepsilon_r)$ by $B_r$ is defined by duality: $\varepsilon_r = *\delta_r$.

Now, we introduce the notion of simple point witch is fundamental for the definition of topology preserving transformations in discrete spaces. We start by giving a definition of local characterization of simple points in $E = \mathbb{Z}^2$. Let consider two neighbourhoods relations $\Gamma_4$ and $\Gamma_8$ defined by, for each point $x \in E$:
$$\Gamma_4(x) = \{y \in E; |y_1 - x_1| + |y_2 - x_2| \leq 1\},$$
$$\Gamma_8(x) = \{y \in E; \max |y_1 - x_1|, |y_2 - x_2| \leq 1\}.$$
In the following we will denote by $n$ a number such that $n = 4$ or $n = 8$ thus we define:
$$\Gamma_n^*(x) = \Gamma_n(x) \setminus \{x\}.$$

We say that a point $y \in E$ is $n-adjacent$ to $x \in E$ if $y \in \Gamma_n^*(x)$. Thus we can generalize for two points $x$, $y$ belonging to $X$. They are n-connected in $X$ if there is a n-path in $X$ between these two points.

The equivalence classes for this relation are the n-connected components of $X$. A subset $X$ of $E$ is said to be n-connected if it consists of exactly one n-connected component. The set composed of all n-connected components of $X$ which are n-adjacent to a point $x$ is denoted by $C_n[x, X]$. In order to have a correspondence between the topology of $X$ and the topology of $\overline{X}$, we use 'n-adjacency' for $X$ and '$\overline{n}$-adjacency' for $\overline{X}$ with $(n, \overline{n})$ equal to (8; 4) or (4; 8).

Informally, a simple point $p$ of a discrete object $X$ is a point which is 'inessential' to the topology of $X$. In other words, we can remove the point $p$ from $X$ without changing the topology of $X$.

The point $x \in X$ is said simple (for $X$) if each n-component of $X$ contains exactly one n-component of $X \setminus \{x\}$ and if each $\overline{n}$-component of $\overline{X} \cup \{x\}$ contains exactly one $\overline{n}$-component of $\overline{X}$. Let $X \subset E$ and $x \in E$, the two connectivity numbers are defined as follows ($\#X$ refers to cardinality of $X$):
$$T(x, X) = \#C_n\left[x, \Gamma_8^*(x) \cap X\right];$$
$$\overline{T}(x, X) = \#C_{\overline{n}}\left[x, \Gamma_8^*(x) \cap \overline{X}\right].$$
The following property allows us to locally characterize simple points [6][9] hence to implement efficiently topology preserving operators: $x \in E$ is simple for $X \subseteq E \leftrightarrow T(x, X) = 1$ and $\overline{T}(x, X) = 1$

The homotopic alternating sequential filter is a composition of homotopic cuttings and fillings by balls of increasing radius. It takes an original image $X$ and a control image $C$ as input, and smoothes $X$ while respecting the topology of $X$ and the geometrical constraints implicitly represented by $C$, simple illustration is given by figure 1. Based on this filter, authors [5] introduce a general smoothing procedure with a single parameter which allows controlling the degree of smoothing.

Let $X$ be any finite subset of $E$, let $C \subseteq X$, $r \in \mathbb{N}$ and $D \subseteq \overline{X}$. The homotopic alternating sequential filter $(HASF)$ of order $n$ with constraint sets $C$; $D$ is defined as follows:

$$HASF_n^{C,D} = HF_n^D \circ HC_n^C \circ ... HF_1^D \circ HC_1^C$$

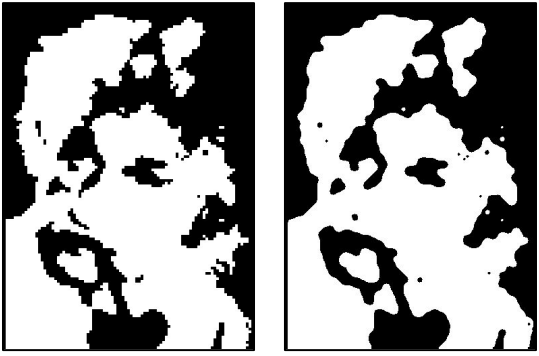
**Fig.1**. (A) Input image (B) Smoothed image

In the previous formula, $HC_n^C$ (**i**) refers to the homotopic cutting of $X$ by $B_n$ with a constraint set $C$. $HF_n^D$ (**ii**) refers to the homotopic filling of $X$ by $B_n$ with a constraint set $D$. These two homotopic operators can be defined as follows:

$$HC_n^C(X) = *H(Y,V) \text{ With } \begin{cases} Y=H(X,\varepsilon_n(X)\cup C) \\ V=(\delta_n(Y)\cap X) \end{cases} \text{ (i)}$$

$$HF_n^D(X) = H(Z,W) \text{ With } \begin{cases} Z=*H(X,\delta_n(X)\cap \overline{D}) \\ W=(\varepsilon_n(Y)\cup X) \end{cases} \text{ (ii)}$$

We recall that $H(Z,W)$ is an homotopic constrained thinning operator. It gives the ultimate skeleton of $Z$ constrained by $W$. Ultimate skeleton is obtained by selecting simple point in increasing order of their distance to the background thanks to a pre-computed quasi-Euclidian distance map [10].

$*H(Y,V)$ is an homotopic constrained thickening operator. It thickens the set of $Y$ by iterative addition of point witch are simple for $\overline{Y}$ and witch belong to the set $V$ until stability.

## 3  Parallelization Strategy

In this section, we start by defining the class of topological algorithms. Then we present our motivation to parallelize it on parallel shared memory machines. Finally we will introduce different steps of our parallelization strategy. We will focus especially on distribution phase and tasks scheduling over different processors.

### 3.1  Class of topological algorithms

In 1996, Bertrand and Couprie [11] introduced connectivity numbers for grayscale image. These numbers describe locally (in a neighborhood of 3x3) the topology of a point. According to this description any point can be characterized following its topological characteristics. They also introduced some elementary operations able to modify gray level of a point without modifying image topology. These elementary operations of point characterization present the fundamental link of large class of topological operators including, mainly, skeletonization and crest restoring algorithms [12]. This class can also be extended, under condition, to homotopic kernel and leveling kernel transformation [13], topological watershed algorithm [14] and topological 2D object smoothing algorithm [5] which is the subject of this article. All mentioned algorithms get also many algorithmic structure similarities. In fact associated characterizations procedures evolve until stability with induce common recursively between different algorithms. Also the grey level of any point can be lowered or enhanced more than once. Finally, all the mentioned algorithms get a pixel's array as input and output data structure. It is important to mention that, to date, this class has not been efficiently parallelized like other classes as connected filter of morphological operator which recently has been parallelized in Wilkinson's work [15]. Parallelization strategy proposed by Sienstra [16] for local operators and point to point operators can also be cited as example. Hence the need of a common parallelization strategy for topological operators that offers an adapted algorithm structure design space. Chosen algorithm structure patterns that will be used in the design must be suitable for SMP machines.

In reality, although the cost of communication (Memory-processor and inter-processors) is high enough, shared memory architectures meet our needs for different reasons: (i) These architectures have the advantage of allowing immediate sharing of data with is very helpful in the conception of any parallelization strategy (ii) They are non-dedicated architecture using standard component (processor, memory ...) so economically reliable (iii) They also offer some flexibility of use in many application areas, particular image processing.

## 3.2 Split Distribute and Merge Strategy

In practice the most effective parallel algorithm design might make use of multiple algorithm structures thus proposed strategy is a combination of the divide and conquer pattern and event-based coordination pattern hence the name that we have assigned: SD&M (Split Distribute and Merge) strategy. Not to be confused with the famous approach of mixed-parallelism (combining data-parallelism and task-parallelism), it is important to mention that our strategy (i) represents the last stitch in the decomposition chain of algorithm design patterns and it provides a fine-grained description of topological operators parallelization while mixed-parallelism strategy provides a coarse-grained description without specifying target algorithm. (ii) It covers only the case of recursive algorithms, while mixed-parallelization strategy is effective only in the linear case. (iii) It is especially designed for shared memory architecture with uniform access.

### 3.2.1 Split phase

The Divide and Conquer pattern is applied first by recursively breaking down a problem into two or more sub-problems of the same type, until these become simple enough to be solved directly. Splitting the original problem take into account, in addition to the original algorithm's characteristics (mainly topology preservation), the mechanisms by which data are generated, stored, transmitted over networks (processor-processor or memory-processor), and passed between different stages of computation.

### 3.2.2 Distribute phase

Work distribution is a fundamental step to assure a perfect exploitation of multi-cores architecture's potential. We'll start by recalling briefly some basic notion of distribution techniques then we introduce our minimal distribution approach that is particularly suitable for topological recursive algorithms where simple point characterization is necessary. Our approach is general and applicable to shared memory parallel machines.

In effect, non-real-time system scheduler doesn't know, in advance, necessary time to perform each task. Thus, "Symmetric Multiprocessing" scheduler distributes tasks to minimize total execution time without load balancing between processors. We propose a novel tasks scheduling approach to prevent improper load distribution while improving total execution time. In literature, there are several schedulers that provide a balanced distribution of tasks such as RSDL "Rotating Staircase Deadline" [17] and CFS "Completely Fair Scheduler" [18]. These schedulers are based on tasks uniformity principle. Through the tasks homogeneity, better distribution can be achieved and total execution time reduced. Unfortunately, these schedulers are not available in all operating system versions. Based on the same principle of tasks uniformity, we propose a new scheduler, simpler to implement and more adapted to topological algorithm processing.

Let be a non-preemptive scheduler 'NPS', $T$ is the set of all tasks, $T_T$ is the set of tasks to process with $T_T \subset T$, $P$ is the set of all processors and $P_D$ is the set of available processors with $P \subset P_D$.

We define 'NPS', $T_x \Rightarrow P_y$, as the scheduler of $T_x$ tasks on $P_y$ processor and $\{p\}$ the increase of $p$.

If $([P_D \neq \varnothing] \wedge [T_T \neq \varnothing])$ then $T_x \Rightarrow P_y : T_x \in T_T$; $P_y \in P_D$. In this scheduler, each processor will treat at maximum $m = \left\{ \dfrac{|T|}{|P|} \right\}$ tasks. Let's consider the following equation, with initial value $\max_0(X) = 0$:
$\max_n(X) = \max(X \setminus \max_{n-1}(X))$.

Then, the worst case to process $T$ will be:
$K(T) = \{\max_1(T), \max_2(T), ..., \max_m(T)\}$.

To demonstrate that let suppose that exist a set $L(T)$ as $\sum L(T) \geq \sum K(T)$. As 'NPS' manage $L(T)$ and $K(T)$, so we can introduce the following: $|L(T)| \leq m$ and $|K(T)| \leq m$.

Now, If $(\sum L(T) \geq \sum K(T))$ then it exist at least one task $\{l\}$, with $k \in K(T)$, such as :
$(l \in L(T)) \wedge (l \notin K(T)) \wedge (l > k)$.
This is impossible according to the definition of $K(T)$. We remember that $K(T)$ was defined as the worst case.

### 3.2.3 Merging phase

The key problem of each parallelization is merging obtained results. Normally this phase is done at the end of the process when all results are returned by all threads what usually means that only one output variable is declared and shared between all threads. In the case of topological operators, we are dealing with a dynamic evolution process so we can plan the following: since two threads finished, they directly merge and a new thread is created. In thread's merging, there is no hierarchical order, the only criteria is finish time.

## 4 Parallel smoothing filter

In this section we start by analyzing overall structure of the original algorithm. Then we move to the parallelization of the Euclidean distance algorithm, thinning algorithm and thickening algorithm. We conclude by a performance analysis of the entire smoothing topological operator.

As we have shown in Section 2, the algorithm receives as input a binary image and the maximum radius. It uses two procedures for homotopic opening and closing. The call is looped to ensure an ongoing relationship between input and output. The opening process is a consecutive execution of erosion, thinning, dilatation and thickening. The closure procedure ensures the same performance of the four consecutive functions with a single difference: the erosion instead of dilatation.

|  | 200x200 ||| 168x288 |||
|---|---|---|---|---|---|---|
|  | r=5 | r=10 | r=∞ | r=5 | r=10 | r=∞ |
| **EucDis** (%) | 64.44 | 54.93 | 46.67 | 59.25 | 49.79 | 35.25 |
| **TopCar** (%) | 8.89 | 13.89 | 18.15 | 11.58 | 16.50 | 24.03 |

**Tab.1.** Time execution rate

Thinning and thickening ensure the topological control of erosion and dilatation. This control is based on researching and removing all destructible points. When a point is deleted, these neighbors are reviewed to ensure that they are not destructible, either. A preliminary assessment of the code, see Table 1, shows that Euclidean distance computing (**EucDis**) takes more time than topological point characterization (**Topcar**). For an image of 200x200, the computation time of the Euclidean distance with an infinite radius is 46.67% while point characterization of 2.4 million points occupies only 18.15%.

If we limit the radius between 5 and 10, the computation time of the Euclidean distance continues to increase. It can reach 64.44% of total time with a radius equal to 5. However time for topological characterization is only 8.89% for 1 million points.

### 4.1 Euclidean distance computing

During previous evaluation, 4SED [10] algorithm was used for Euclidean distance computation. So we are looking for another algorithm that is faster, and parallelizable. The new algorithm must have an Euclidean distance computation's error less than, or equal to, that produced by 4SED in order to maintain homotopic characteristics of the image.

In literature, several algorithms for Euclidean distance computing exist. Lemire [19] and Shih [20] algorithms are bad candidates because Lemire's algorithm does not use Euclidean circle as structuring element. Then homotopic property will not be preserved. Shih and al. algorithm has strong data dependency which penalizes parallelization. In [21], Cuissenaire propose a first algorithm, called PSN "Propagation Using a Single Neighborhood" that uses the following structure element:

$$d_4(p) = \left\{ q \vee \sqrt{(q_x - p_x)^2 + (q_y - p_x)^2} < 1 \right\}.$$

He also proposes a second algorithm, called PMN "Propagation Using Multiple Neighborhood", which uses the eight neighbors. In [22], Cuissenaire proposes a third algorithm with $o(n^{3/2})$ complexity, which offers an accurate computation of the Euclidean distance. The only drawback of this algorithm is computation time witch is very important and goes beyond the two algorithms mentioned above. Even if computing error produced by PSN is greater than computing error produced by PMN, it is comparable to that produced by 4SED. The low data dependence and the ability to operate on 3D images, makes PSN algorithm a potential candidate for replacing 4SED. Meijster [23] proposes an algorithm to compute exact Euclidean distance. The algorithm complexity is $o(n)$ and it operates in two independent, but successive, steps. First step is based on looking over columns then computing distance between each point and existing objects. Second step includes same treatment for lines. It is important to note that strong independence between different processing steps and computing error equal to zero makes Meijster algorithm a potential candidate to replace 4SED. It is also able to operate on

3D images. Theory analysis of Meijster and Cuissenaire algorithms can be found in Fabbri's work [24].

In the following, we propose a first analysis based on different algorithms implementation in order to compare between them. We have implemented 4SED algorithm using a fixed size stack. It uses a FIFO queue and it has a small size while 4SED algorithm does not need to store temporal image. Results are directly stored into the output image. We will retain this implementation because 4SED assessment serves only as reference for comparison. For PSN implementation, we used stacks with dynamic sizes. The memory is allocated using small blocks defined at stack creation. When an object is added to the queue, the algorithm will use the available memory of the last block. If no space is available, a new block is allocated automatically. Block size is proportional to the image size (N x M / 100). Finally we used a simple memory structure for the implementation of Meijster algorithm. A simple matrix was used to compute distance between points and object of each column. Three vectors were used to compute distance in each line. Figure 2 describes obtained results by the different three implementations. During this evaluation we used a binary test image (200x200). We have also modified ball radius. We used Valgrind software to evaluate different designs. Callgrind tool returns the cost of implementing of each program by detecting IF (Instruction Fetch).

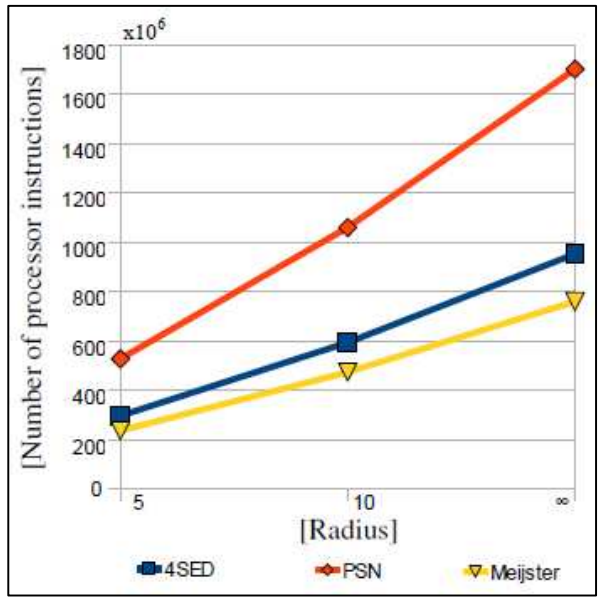

**Fig.2.** Overall structure of the original algorithm

Results show that PSN algorithm is the most expensive in all cases (for any radius). Meijster algorithm is moderately 5x faster than 4SED. The output images returned by Meijster algorithm hold the best visual quality while Euclidean distance computation error is almost zero. So our efforts will be brought on Meijster algorithm parallelization.

We denote by $I$ the input image with $m$ columns and $n$ rows. We denote by $B$ an object included in $I$. The idea is to compute, for each point $p \in I \land p \notin B$, the separating distance between $p$ and the closest point $b$ with $b \in B$. This amounts to compute the following matrix: $dt[p_x, p_y] = \sqrt{EDT(p)}$ With $EDT(p) = \min(p_y - b_x)^2 + G(p_x, b_y)^2$ and, $\forall (0 \leq b \leq m)$, $b = (b_x, b_y)$. If we assume that minimum distance of an empty group $K$ is $\infty$ and $\forall z \in K$, we have $(z_y + \infty) = \infty$ then $EDT(p)$ formula can be written as follow: $\forall b_x < n$, $\forall b_y \leq m$,

$EDT(p) = \min(p_y - b_x)^2 + G(p_x, b_y)^2$ with

$G(p_x, y) = \min |p_x - b_x| : b = (b_x, y)$.

Thus we can split the Euclidian distance transform procedure into two stages. The first step is to scan columns and compute $EDT$ for each column $y$. The second step consists on repeating the same procedure for each line. In the following we start by detailing these two steps: In the first step $G(p_x, y)$ can be computed through the two following sub functions:

$G_T(p_x, y) = \min p_x - b_x : b = (b_x, y) \forall 0 \leq b_x \leq n$
$G_B(p_x, y) = \min b_x - p_x : b = (b_x, y) \forall 0 \leq b_x \leq n$

To compute $G_T(p_x, y)$ and $G_B(p_x, y)$, we scan each column $y$ from top to bottom using the two following formula:

$G_T(p_x, y) = G_T(y, p_x - 1) + 1$
$G_B(p_x, y) = G_B(y, p_x + 1) + 1$.

Let's move to the second step, we start by defining $f(p, y) = (p_y - y)^2 + G(p_x, y)^2$. Then we can define $EDT(p) = \min f(p - y), \forall 0 \leq y \leq m$. For each row $u$, we note that there is, for the same point $p$, the same value of $f(p, y)$ even if $y$ change its values, so we can introduce the concept of "region of column ".

Let $S$ be the set of all $y$ points such that $f(p, y)$ is minimal and unique. The formula of $S$, $\forall 0 \leq y \leq u$, will be $S_p(u) = \min y : f(p, y) \leq f(p, i)$.

Let $T$ be the set of all points having coordinates greater than, or equal to, horizontal coordinate of the intersection with a region. The formula of $T$, $\forall((0 \leq i \leq u) \wedge (u \leq m))$, will be the following :
$$T_p(u) = Sep_{p_x}(S_p(u-1), u) + 1$$

Let $Sep(i, u)$ be the separation between regions of $i$ and $u$, defined by:

$$f(p, i) \leq f(p, u)$$
$$\Leftrightarrow (p_y - i)^2 + G(p_x, i)^2 \leq (p_y - u)^2 + G(p_x, u)^2$$
$$\Leftrightarrow Sep_{p_x}(i, u) = (u^2 - i^2 + Dif) / 2(u - 1) = p_y$$
With $Dif = (G(p_x, u)^2 - G(p_x, i)^2)$.

Thus lines processing will be, at the beginning, from left to right then from right to left. During the first term, from left to right, two vectors $S$ and $T$ will be created. These two vectors will contain respectively all regions and all intersections. During the second treatment, from right to left, we compute $f$ for each value of $S$. $f$ is computed also for each respective values of $T$.

The independence of data processing between rows and columns is the key to apply SDM parallelization strategy. In the first stage, column processing, we can define data interdependence by the following equation:

$$G(p_x, y) = \min\{G_T(p_x, y), G_B(p_x, y)\};$$
$$\Leftrightarrow G_T(p_x, y) = \begin{Bmatrix} 0 & if (p_x, y) \in B \\ G_T(p_x, y) & else \end{Bmatrix};$$
$$\Leftrightarrow G_B(p_x, y) = \min\{G_B(p_x + 1, y), G_T(p_x, y)\}.$$

It follows that values of each column y of G, depends only on lines: $p_x$, $p_x + 1$ and $p_x - 1$. Similarly, at the second stage, we can introduce the following interrelationship: $Edt(p) = f(p, S_p(q))$.

Then $\forall (0 \leq y \leq u), (0 \leq i \leq u) \wedge (u < m)$,
$S_p(u) = \min y : f(p, y) \leq f(p, i)$ ;
Thus, if $(u = T_p(q))$ so $q = (q - 1)$ which imply the following: $T_p(u) = Sep_{p_x}(S_p(q), u) + 1$.

According to this formalization, values of $f(p, i)$ and $Sep_x(i, u)$ are independent of modified data. So using two vectors $S$ and $T$, a private variable $q$ for each line ensures complete independence in writing.

We start applying the splitting step by sharing the columns and lines processing between multiple processors. A thread can process one or more columns and the number of threads used will depend on the number of processors. The results returned by all threads in this first stage will be merged in order to start lines processing. In the following we introduce the parallel version of the algorithm Meisjter for both steps. Associated algorithm complexity is $o((n \times m) / N)$. $(n \times m)$ refers to image size and $N$ refers to the number of processors.

| Algorithm 1: Parallel Version Meijster [1st step] |
|---|
| 1.    **For** $(y = t, y < m, y = y + t_{max})$ **do** |
| 2.       **If** $(0, y) \in B$ **then** $g[0, y] \leftarrow 0$ |
| 3.            **else** $g[0, y] \leftarrow \infty$ |
| 4.       **endif** |
| 5.       */\* $G_T$ \*/* |
| 6.       **for** $(x = 1)$ **to** $(n - 1)$ **do** |
| 7.          **if** $[x, y] \in B$ **then** $g[x, y] \leftarrow 0$ |
| 8.          **else** $g[x, y] \leftarrow g[x + 1, y] + 1$ |
| 9.          **endif** |
| 10.      **endfor** |
| 11.      */\* $G_B$ \*/* |
| 12.      **for** $(x = n - 2)$ **downto** $(0)$ **do** |
| 13.          **if** $(g[x + 1, y] < g[x, y])$ **then** |
| 14.             $g[x, y] \leftarrow g[x + 1, y] + 1$ |
| 15.          **endif** |
| 16.      **endfor** |
| 17. **endforall** |

**Algorithm 2: Parallel Version Meijster [2$^{nd}$ Step]**

1.    For $(x = t, x < n, x = x + t_{max})$ do
2.       $q = 0$
3.       $s[0] = 0$
4.       $t[0] = 0$
5.       */\* First part \*/*
6.       for $(u = 1)$ to $(m - 1)$ do
7.          $A \leftarrow (q \geq 0) \wedge \left[ f((x, t[q]), s[q]) \right]$
8.          $B \leftarrow f((x, t[q]), u)$
9.          while $(A > B)$ do $q \leftarrow (q + 1)$
10.      end while
11.
12.      if $(q < 0)$ then $(q \leftarrow 0)$
13.              $(s[0] \leftarrow u)$
14.      else $w \leftarrow Sep(s[q], u, x) + 1$
15.          if $(w < m)$ then $q \leftarrow (q + 1)$
16.              $s[q] \leftarrow u$
17.              $t[q] \leftarrow w$
18.         endif
19.       endif
20.      endfor
21.      */\* Second part \*/*
22.      for $(u = m - 1)$ downto $(0)$ do
23.         $Edt[x, u] \leftarrow f((x, u), s[q])$
24.         if $(u = t[q])$ then $q \leftarrow (q - 1)$
25.         endif
26.      endfor
27.   end forall

### 4.2 Thinning and thickening computing

Algorithms of thinning and thickening are almost the same. The only difference between them is the following: in thinning algorithm, destructible points are detected then their values are lowered. In thickening algorithm, constructible points, are detected then their values are increased. For parallelization, we will apply the same techniques introduced in [25]. We propose a similar version using two loops. Target points are initially detected then their value lowered or enhanced according to appropriate treatment. The set of their eight neighbors are copied into a temporary "buffer" and rechecked. This process is repeated until stability. In the following, we present an adapted version of Couprie's thinning algorithm.

**Algorithm 3: Adapted Version Thinning Algo.**

1.   while $(input[x]$ is destructible$)$ do
2.      $push(x, stack1)$
3.      $x \leftarrow x + 1$
4.   endwhile
5.   $output \leftarrow input$
6.   While $(stack1 \neq \varnothing) \wedge (\max_{iter} > 0)$ do
7.      While $(stack1 \neq \varnothing)$ do
8.         $x \leftarrow pop(stack1)$
9.         if $(output[x]$ is destructible$)$ then
10.           $output[x] \leftarrow reduce\_pt(x)$
11.           $push(x, stack2)$
12.         endif
13.      end while
14.      While $(stack2 \neq \varnothing)$ do
15.         $x \leftarrow pop(stack2)$
16.         $v \leftarrow neighbors(x)$
17.         $i \leftarrow 0$
18.         While $(i < 8)$ do
19.           if $(v[i] \notin stack1)$ then
20.             $push(v[i], stack1)$
21.           endif
22.         endwhile
23.      endwhile
24.      $\max_{iter} \leftarrow \max_{iter} - 1$
25.   endwhile

### 4.3 Experimental analyses

Proposed parallel version of topological smoothing algorithm was implemented in C using OpenMP directives. We implemented two versions, the first one using 'Symmetric Multiprocessing' scheduler and the second one using 'non-preemptive' scheduler. Wall-clock execution times for numbers of threads equal to 1, 2, 4, 8, and 16 were determined. The efficiency measure $\Psi(n)$ is given by the following formula with $n$ the number of processors: $\Psi(n) = t_s / (n * t_p)$, $t_s$ refers to the serial time and $t_p$ refers to parallel time. Times were performed on eight-core (2× Xeon E5405) shared memory parallel computer, on Intel Quad-core Xeon E5335, on Intel Core 2 Duo E8400 and Intel mono-processor Pentium 4 660. Each processor of the Xeon E5405 and E5335 runs at 2 GHz and both of the two machines have 4 GB of RAM. The E8400 processor runs at 3GHz. The Pentium processor runs at 3.6 GHz. The last two machines have 2 GB of RAM. The minimum value of 5 timings was taken as most indicative of algorithm speed. The measurements were done on 2D binary image (512x512). Results of the second implementation, using non preemptive schedule, are shown in the following figure :

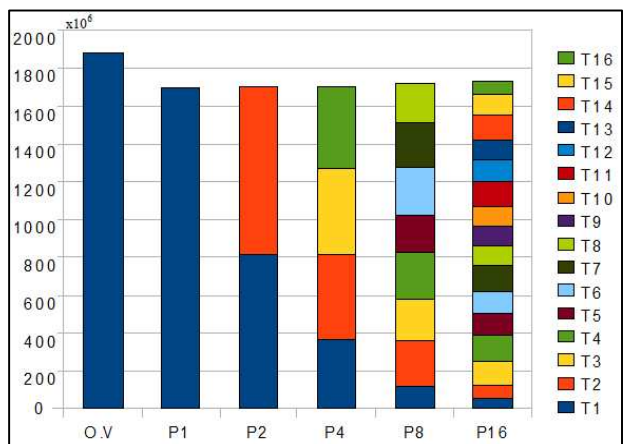

**FIG. 3.** Number of Instructions and tasks distribution using 'NPS'

On the eight-core machine, Number of instructions in the second implementation drops from an average of 1879 x10$^6$ FI with a single thread down to 202x10$^6$ ms with 8 threads. As expected, the speed-up for the second implementation using 'non preemptive' scheduler is higher than for the one using "Symmetric Multiprocessing" scheduler, thanks to balanced distribution of tasks. A remarkable result, shown in figure 4, is the fact that speed-up increases as we increase the number of threads beyond the number of processors in our machine (eight cores). In a first implementation, using "Symmetric Multiprocessing" scheduler, the speedup at 8 threads is 1.9 ± 0.01. However, for the second implementation, using our scheduler, the speedup has increased to 5.2 ± 0.01. Thus execution time is decreased from 0.16± 0.005s to 0.03± 0.001s. This performance has allowed a cadency of 32 images per second (see Figure 5). This real-time performance confirms the interest of proposed parallelization strategy. Another common result between different architecture is stability of execution time on each n-core machine since the code uses n or more threads (see figure 4).

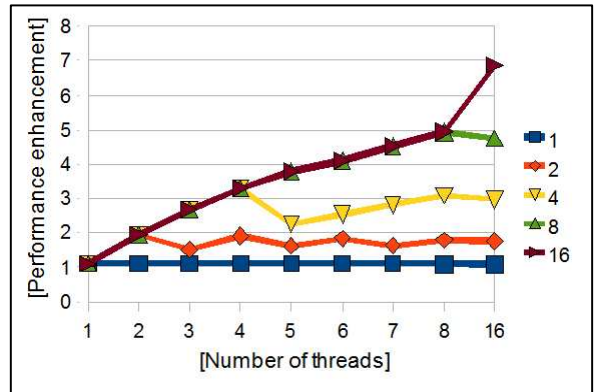

**FIG. 4.** Performance improvement using 'NPS'.

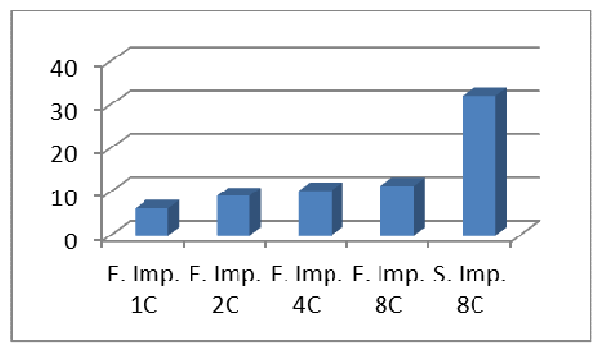

**FIG. 5.** Summary of performance in term of cadency: (F. Imp. NC = First Implementation using Symmetric Multiprocessing scheduler on N processors. S. Imp. NC = Second Implementation using Non Preemptive scheduler on N processors)

For better readability of our results, we tested the efficiency of our algorithm on various architectures (see figure 5) using the $\Psi(n)$ formula introduced earlier with $t_s$ sequential time on mono-processor Pentium 4 660. For parallel time we use best parallel time obtained using 8 threads with Non Preemptive scheduler.

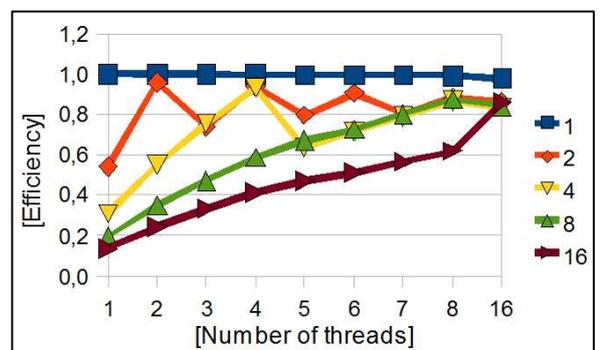

**FIG. 6.** Efficiency improvement

## 5 Conclusion

Topological characteristics are fundamental attributes of an object. In many applications, it is mandatory to preserve or control the topology of an image. Nevertheless, the design of transformations which preserve both topological and geometrical features of images is not an obvious task, especially for real time processing.

In this paper, we have presented a new parallel computation method for topological smoothing. We have also presented an adapted parallelization strategy. SDM-strategy was a conditional application of the well known principle of divide and conquers associated to event-based coordination techniques. First major contribution in this paper is the parallel computation method for image smoothing allowing real-time processing with topology preservation. Second contribution is the non-specific nature of proposed parallelization strategy. In fact it can be applied for a large class of topological operators as we shown in section 3.1. Third contribution concern tasks distributions. We presented a non-preemptive scheduler 'NPS', simpler to implement and more adapted to particular topological algorithm processing.

Parallel computation of topological operators represents many challenges, ranging from parallelization strategies to implementation techniques. We tackle these challenges using successive refinement, starting with highly local operators, which process only by characterizing points and then deleting target pixels, and gradually moving to more complex topological operators with non-local behavior. In future work, we will study parallel computation of the topological watershed [14].